# Au$_4$Mn, a localized ferromagnet with strong spin-orbit coupling, long-range ferromagnetic exchange and high Curie temperature.

Yangkun He[1,2,*], Zsolt Gercsi[1], Rui Zhang[1], Yu Kang[2], Yurii Skourski[3], Lucy Prendeville[1], Orrie Larmour[1], Jean Besbas[1], Claudia Felser[2], Plamen Stamenov[1] and  J. M. D. Coey[1]

*1. School of Physics and CRANN, Trinity College, Dublin 2, Ireland*

*2. Max-Planck-Institute for Chemical Physics of Solids, 01187 Dresden, Germany*

*3. Dresden High Magnetic Field Laboratory (HLD-EMFL), Helmholtz-zentrum Dresden–Rossendorf, Dresden 01328, Germany*

E-mail: heya@tcd.ie (Dr. Yangkun He)

**Abstract**

Metallic Mn-based alloys with a nearest-neighbor Mn-Mn distance greater than 0.4 nm exhibit large, well-localized magnetic moments. Here we investigate the magnetism of tetragonal Au$_4$Mn with a Curie temperature of 385 K, where manganese has a spin moment of 4.1 $\mu_B$ and its orbital moment is quenched. Since 80% of the atoms are gold, the spin orbit interaction is strong and Au$_4$Mn exhibits uniaxial magnetocrystalline anisotropy with surface maze domains at room temperature. The magnetic hardness parameter of 1.0 is sufficient to maintain the magnetization along the *c*-axis for a sample of any shape. Au also reduces the spin moment of Mn through 5*d*-3*d* orbital hybridization. An induced moment of 0.05 $\mu_B$ was found on Au under a pulsed field of 40 T. Density functional theory calculations indicate that the Mn-Mn exchange is mediated by spin-polarized gold 5*d* and 6*p* electrons. The distance-dependence shows that it is ferromagnetic or zero for the first ten shells of Mn neighbors out to 1.041 nm (64 atoms), and very weak and oscillatory thereafter.

**Introduction**

The magnetism of Mn depends critically on the interatomic manganese distance, giving a rich variety of possible magnetic structures. Generally, manganese has no magnetic moment for the shortest Mn-Mn distances (*d* < 0.24 nm); then a moment develops and the exchange coupling is antiferromagnetic for more typical nearest neighbor separations (0.25 nm < *d* < 0.28 nm); bigger atomic moments and ferromagnetic coupling is found at longer separations (*d* > 0.29 nm) [1]. Although the general rule of ferromagnetism of Mn in ordered alloys with long Mn-Mn distances is known, the type of exchange interaction and magnetism (localized or itinerant) is not clear. For instance, dilute alloys such as **Cu**Mn with the largest Mn-Mn separations are spin glasses.

Au$_4$Mn crystallizes in Ni$_4$Mo-type tetragonal *I*4/m structure with Au occupying 8h sites (0.2, 0.4, 0) and Mn on 2*a* sites (0, 0, 0) [2] as shown in Fig. 1a. Each Mn atom has twelve Au neighbors at almost the same distance of ~0.287 nm. Each Mn aligns with its two Mn nearest neighbors (1$^{st}$ NN) in chains along the *c* axis (*d* is 0.404 nm) and there are eight next nearest neighbors (2$^{nd}$ NN) at 0.5 nm in the four surrounding chains, which also couple ferromagnetically. Localized manganese moments of ~4.1 $\mu_B$ that couple ferromagnetically along the *c* axis were reported in early work [2]. The material is prototypical to study the magnetism of Mn at large Mn-Mn separations. What makes this material also interesting is its high Curie temperature $T_C$ (~380 K) [3] in an almost dilute system with only 20 atomic % Mn. In addition, in the Au$_4$X structure, X can be either an early 3*d* transition element or a late 4f element [4]. Thus, it is a rare example of a system where the magnetism of 3d and 4f elements can be compared in the same crystal field environment.

In the light of the above, the magnetism of Au$_4$Mn raises some questions that we address in this paper: (1) Is the magnetocrystalline anisotropy uniaxial and strong enough to resist the demagnetizing field for a sample of any shape? What is the domain structure? (2) What is the influence of spin-orbit coupling? Is the orbital moment of Mn quenched, or does it contribute to the total moment as in a rare earth? (3) What is the nature of the exchange interactions $\mathcal{J}$, direct or RKKY? (4) What does the band structure look like? Is Au magnetic in Au$_4$Mn? Here we address these questions experimentally, and compare the results with density functional theory calculations for Au$_4$Mn.

**Methods**

High-purity Au and Mn were mixed in an Al$_2$O$_3$ crucible before sealing them in a quartz tube under Ar. The tube was heated at 1100 °C for one day for homogeneity. It was then slowly cooled to 350 °C over five days, followed by a three-week anneal to improve the atomic order of the polycrystalline material. The crystal structure was characterized by X-ray diffraction (XRD). Magnetization was measured in a superconducting quantum interference device magnetometer (SQUID, Quantum Design). High-field magnetization measurements were performed in the Dresden High Magnetic Field Laboratory in pulsed fields of up to 56 T. The magnetic domain images were observed by a magneto-optical Kerr microscope (Evico Magnetics) at room temperature on a polished surface. A rare-earth iron garnet film (with saturation field 0.8 mT) was used to reveal the stray field at the surface. Ferromagnetic resonance measurements (FMR) were performed using a Bruker EMX EPR spectrometer with an X-band bridge operating at 9.614 GHz on a plate-like sample with the field perpendicular to the plane. The static magnetic field was ramped from 0 to 0.6 T and the field derivative of the absorbed microwave power was measured as a function of the magnetic field. *Ab-initio* calculations based on density functional theory (DFT) were carried out using norm-conserving pseudopotentials and pseudo-atomic localized basis functions implemented in the OpenMX software package [5]. The generalized gradient approximation GGA-PBE [6] was used for all the calculations. The structure was fully relaxed to minimize interatomic forces. We used a 10-atom superstructure with 8 Au and 2 Mn atoms with (11×11 ×15) *k*-points to evaluate the total energies. This extended structure also allowed us to simulate the electronic structure with both ferromagnetic (FM) and antiferromagnetic (AFM) couplings of the manganese. Pre-generated fully relativistic pseudopotentials and the pseudo-atomic orbitals with a cut-off radius of 6 atomic units (a.u.) were used with s3p3d3 for Mn and 7 a.u. with s3p3d3f1 for Au. An energy cut-off of 300 Ry was used for the numerical integrations. The convergence criterion for the energy minimization procedure was set to $10^{-8}$ Hartree. The spin orbit interaction (SOI) was turned on for the calculations. The magnetocrystalline anisotropy (MAE) was calculated using a 5-atom cell and a large number of k-points (15 × 15 × 19) for high precision.

**Results**
**Crystal structure**

The crystal structure was characterized by X-ray diffraction (XRD) on polycrystalline material. The total time for XRD is 56 hours. Since Au$_4$Mn is not brittle, we were unable to make strain-free powders for diffraction experiment. We used a polished bulk sample and there may be some texture that could influence the relative intensities of the reflections. The crystal structure and the corresponding XRD pattern are shown in Fig. 1. The fitted lattice constants in Ni4Mo-type tetragonal *I*4/*m* structure are $a = 0.6459(8)$ nm and $c = 0.4042(3)$ nm. The crystal has a tetragonally-distorted face-center-cubic structure [2], where the lattice constants are $a' = \sqrt{2/5}a = 0.4085(5)$ nm, $c' = c = 0.4042(3)$ nm, as shown in Fig. 1a. $c'$ is slightly smaller than $a'$ ($c'/a' = 0.99$) because of contraction along the [001] Mn chains. Atomic ordering was confirmed by the superlattice (120) peak and the splitting of the fcc (402) peak.

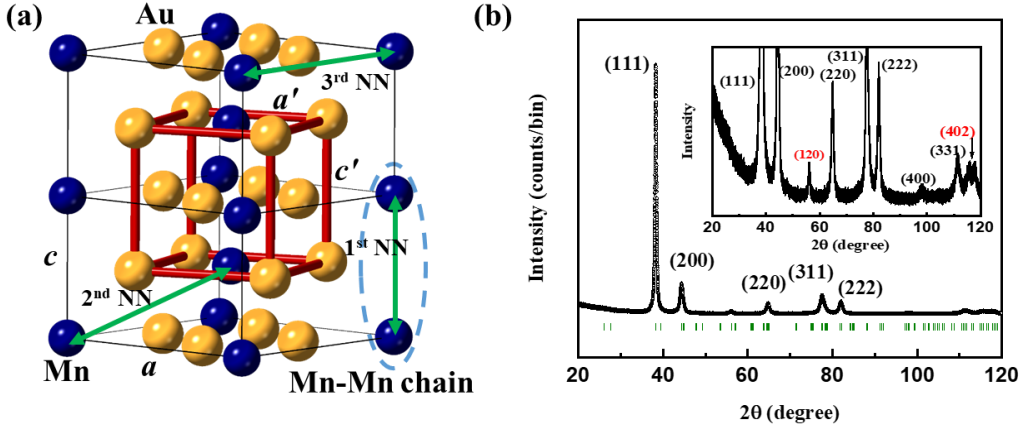

Figure 1. (a) Crystal structure of Au4Mn (black unit cell showing a double cell along *c* axis) as a distorted fcc structure (red cell). Blue and yellow atoms are Mn and Au respectively. The first nearest neighbor (1st NN) to third nearest neighbor (3rd NN) Mn pairs are shown. (b) XRD pattern of Au4Mn. The Bragg peak positions are shown by green vertical bars. The indices shown are based on the fcc structure. The inset shows the enlarged XRD pattern.

**Magnetic properties**

The magnetization curves at different temperatures are shown in Fig. 2a. The extrapolated spontaneous moment $m_s$ in zero field measured at 4 K is 4.1 $\mu_B$/f.u., which decreases to 2.8 $\mu_B$/f.u. at room temperature. The high field slope of the magnetization curve at 4 K is quite steep, 0.02 $\mu_B$/T, which is similar to that of ferromagnetic Au4V with the same structure [7]. This slope is accompanied by a large volume magnetostriction [8]. High field magnetization data shown in Fig. 2b saturates at 40 T with a moment $m = 4.5\pm0.1$ $\mu_B$/f.u. The additional moment of 0.4 $\mu_B$ induced by the high magnetic field is similar to that of Au4V [7]. The spontaneous moment as a function of temperature is plotted in Fig. 2c. Mean field theory was used to fit the curve with S = 2 (S is the spin quantum number), which matches the experiment. The paramagnetic Curie temperature $\theta_p$ fitted by the mean field model is 387 K. The inverse susceptibility $\chi^{-1}$ as a function of temperature is plotted in Fig. 2d, indicating an effective magnetic moment $m_{eff}$ of 4.9 $\mu_B$, in agreement with a previous report [9], and a localized moment with S = 2.

The magnetocrystalline anisotropy was determined by the anisotropy field deduced from ferromagnetic resonance (FMR) at room temperature as shown in Fig. 2e. For an isotropic free electron, the frequency of 9.614 GHz corresponds to a remanence $\mu_0 H_r = 0.343$ T. In our polycrystalline sample the total signal comes from a distribution of crystallites with slightly different resonance frequencies. Each frequency is determined by the direction of the applied field, the crystallite magnetization and its magnetocrystalline anisotropy. The low field FMR position corresponds to the case when the applied field is parallel or close to the easy axis of the crystals in the polycrystalline sample; the high field position indicates the field is nearly in the *ab*-plane [10,11]. For the low field FMR position if we ignore the high-order anisotropy terms, the relationship between the FMR field position $\mu_0 H_R$, anisotropy field $2K/M_s$ and the angular frequency of the microwave field $\omega$ satisfy [10]

$$\left(\frac{\omega}{\gamma}\right)^2 = \left(\mu_0 H_R + \mu_0 M(N_x - N_z) + \frac{2K_1}{M_s}\right)\left(\mu_0 H_R + \mu_0 M(N_y - N_z) + \frac{2K_1}{M_s}\right) \quad (1)$$

where $\gamma$ is the gyromagnetic ratio of 28 GHz T$^{-1}$ and $N_x$, $N_y$ and $N_z$ are the demagnetizing factor along different directions. For our plate-like sample with the field perpendicular to the plane, $N_x = N_y = 0$ and $N_z = 1$. Using Eq. 1, the anisotropy field $2K_1/\mu_0 M_s$ at room temperature is roughly estimated as 0.76 T. The corresponding value of $K_1$ is 100±10 kJm$^{-3}$, which agrees with a previous torque measurement [2]. The magnetic hardness parameter $\kappa = \sqrt{K_1/(\mu_0 M_s^2)}$, a convenient figure of merit for permanent magnets [12], is 1.0. At low temperature $K_1$ increases to 300 kJm$^{-3}$ [2], while $\kappa$ retains the same value. This means Au$_4$Mn is a highly anisotropic material, which can maintain the perpendicular anisotropy in any shape below $T_C$.

The magnetic domain structure was observed on a polished surface after demagnetization by an AC magnetic field, as shown in Fig. 2f. Maze surface domains, a sign of strong uniaxial anisotropy, were observed, similar to those in many hard magnetic materials such as Nd$_2$Fe$_{14}$B [13], MnBi [14] and Rh$_2$CoSb [15]. This confirms the large magnetocrystalline anisotropy from the FMR measurement. Note that the Kerr rotation angle in our magneto-optical measurement is lower than the resolution limit (< 0.01 degree) using red (1.96 eV) or blue (3.06 eV) light different from the predicted value of ~0.4 degree [16], therefore a rare-earth iron garnet film (the saturation field is 0.8 mT) was needed to reveal the stray field and domain structure at the surface.

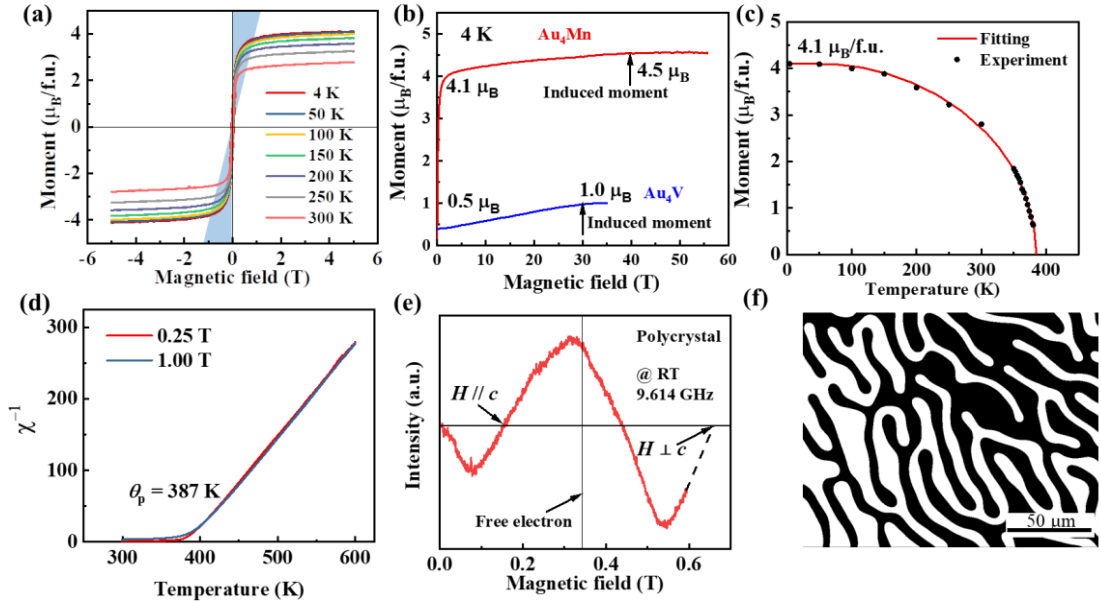

Figure 2. Magnetic properties. (a) Magnetization curves at different temperatures. The multi-domain state exists under a small magnetic field (blue region). (b) Magnetization measured in a pulsed magnetic field of 56 T at 4 K. The saturation was found at 40 T (see the cross of two dashed lines). The data for Au$_4$V are from Ref. [7]. (c) Saturation magnetization as a function of temperature and the S = 2 fit by mean-field theory. The data below 300 K were obtained from the M-H curves; the data close to the $T_C$ were measured under an applied magnetic field of 0.25 T. (d) Inverse susceptibility $\chi^{-1}$ measured under 0.25 T and 1 T. (e) FMR spectrum at room temperature (RT) in the polycrystalline sample. Crystals with their easy axis closer to the field direction exhibit smaller $\mu_0 H_R$ while larger magnetic field is required for crystals with the c axis perpendicular to the field. (f) Magnetic domain structure at room temperature.

**Band structure**

In order to reveal the nature of the exchange interaction and the origin of high $T_C$ in Au$_4$Mn, first principles calculations were used to investigate the band structure. Our calculation shows a similar band structure to a previous study, which discussed the role of Au on magnetism in metallic systems [17]. Fig. 3a shows the total (TDOS) and partial (PDOS) densities of states in Au$_4$Mn. The occupied spin channels in TDOS are symmetrical up to about $E_F$ -2 eV, where the 3d bands of Mn split by onsite exchange dominate for the spin up channel. The unoccupied minority Mn 3d bands are located about 1.5 eV above Fermi level. The magnetic Mn-Mn coupling is strongly ferromagnetic, and with about 20 meV/atom lower in energy as compared to an antiferromagnetic spin alignment. The symmetric spin channels of PDOS for Au suggest a small contribution to the total magnetic moment. The calculated net moment of Au is small, 0.05 $\mu_B$ per atom, in agreement with previous X-ray magnetic circular dichroism (XMCD) measurements [18,19]. The orbital-specific DOS of Mn is shown in Fig. 3b. The contribution to the DOS from 3d bands at the Fermi level is small (0.35 eV$^{-1}$ in the spin down channel shown in Fig. 3c compared to 1-2 eV$^{-1}$ for transition metals Fe, Co and Ni) and free electrons of Au are dominant instead, resulting in a small electronic specific heat of $\gamma$ = 2 mJ/mole K$^2$ [20] and a tiny Kerr rotation angle with visible light.

We compare the DOS of Au 6s, 6p and 5d orbitals in pure fcc Au and Au$_4$Mn in Figs. 3d-3f. There is little difference in 6s and 6p orbital occupations in the spin up channel, but a significant difference is found in the spin down channel close to the Fermi level, which means that the 6s and 6p electrons in Au$_4$Mn are polarized by Mn due to band hybridization. In particular, DOS of Au 6p orbitals shows strong resemblance to Mn 3d orbitals close to the Fermi level as shown in Fig. 3c, indicating its important role in the exchange and the related magnetism of Au$_4$Mn (discussed below). In Au$_4$Mn, the Au 5d bands are symmetric and nearly fully occupied, where the top of the bands is about 2 eV below the Fermi level. A slight shift in Au 5d orbital occupation can be observed due to hybridization with the overlapping majority spin bands of 3d Mn when compared to pure fcc Au.

To further study the influence of the band hybridization on the magnetism, we have calculated the electronic structure of pure Mn in Au$_4$Mn lattice after removing the Au atoms in a hypothetical body-centre-tetragonal (bct) structure. In hypothetical 'bct Mn', the exchange splitting is larger with spin up channel 4 eV below $E_f$ and the moment is a good quantum number of 5.0 $\mu_B$ per Mn, larger than 4.1 $\mu_B$ in Au$_4$Mn with a 3d-5d hybridization.

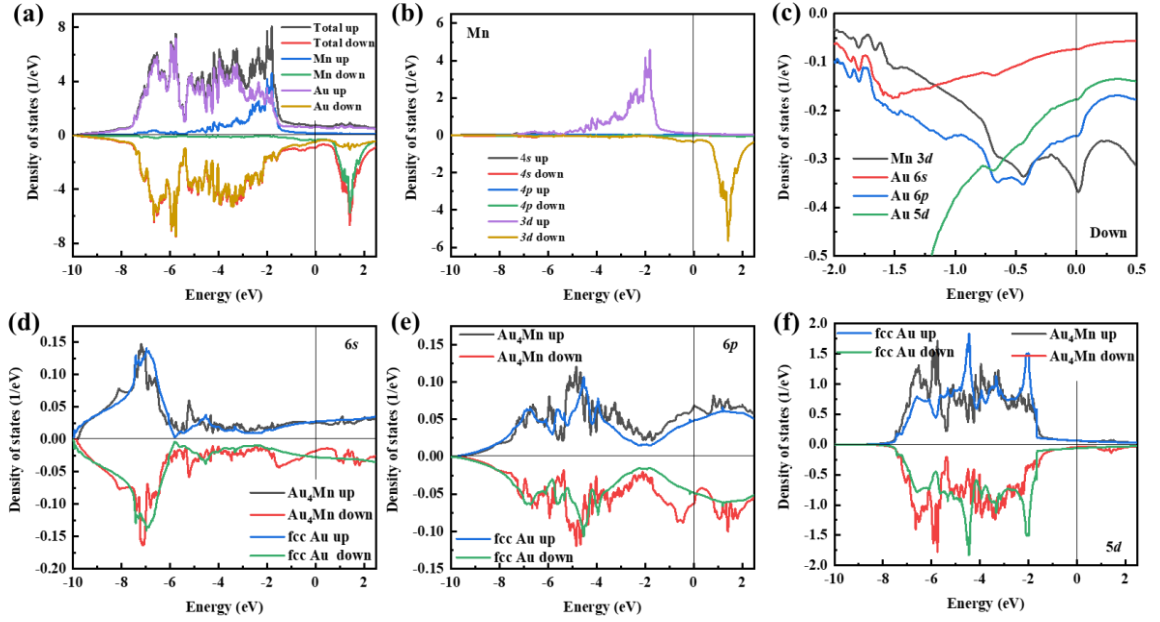

Figure 3. Results of first principles calculations. (a) Partial DOS of Au and Mn in Au$_4$Mn together with total DOS. (b) DOS of Mn 3$d$, 4$s$ and 4$p$ orbitals in Au$_4$Mn. (c) DOS close to the Fermi level in the spin down channel. (d)-(f) Comparison of the DOS of Au 6$s$, 6$p$ and 5$d$ orbitals in pure Au and Au$_4$Mn

**Discussion**

**a. Localized moments**

The series Au$_4$X (X = Sc, Ti, V, Cr, Mn, Er, Tm, Yb, Lu) offers a way to compare the magnetism of 3$d$ and 4$f$ elements sharing the same crystal field environment. We summarize the magnetic properties of the series in Table 1. All these compounds exhibit matching moments from both inverse susceptibility and mean field theory, indicating a localized moment on X. The electronic specific heat of ~2 mJ mol$^{-1}$ K$^{-2}$ [20] in Au$_4$Mn is much lower than itinerant ferromagnets such as CoS$_2$ [21] and Ni [22], indicating a small density of states at the Fermi level from the broad $s$, $p$ band. The tiny Kerr rotation angle in our optical measurement further indicates the absence of any highly spin-polarized band at the Fermi level, similar to Gd [23]. Unlike the rare earths, the orbital moments of the 3$d$ elements are quenched.

Table 1. Magnetic properties of Au$_4$X (X = Sc, Ti, V, Cr, Mn, Er, Tm, Yb, Lu). The $m_s$ and $m_{eff}$ match well for a localized moment. Au$_4$Ti, Sc, Lu are nonmagnetic.

|  | $m_s$ (μ$_B$) | $m_{eff}$ (μ$_B$) | J (Total moment) | | $m_{eff}$ (μ$_B$) | $T_C/T_N$ | Exchange | Ref. |
|---|---|---|---|---|---|---|---|---|
|  | Experiment | Experiment |  |  | Theory | (K) | Interaction |  |
| Au$_4$Mn | 4.1 | 4.9 | S | 2 | 4.9 | 387 | FM | This work |
| Au$_4$Cr | 3.2 | 4 | S | 3/2 | 3.9 | 380 | AFM | [9] |
| Au$_4$V | 1 | 1.7 | S | 1/2 | 1.7 | 57 | FM | [7] |
| Au$_4$Ti | 0 | 0 | - | 0 | 0 | - | - | [3] |
| Au$_4$Sc | 0 | 0 | - | 0 | 0 | - | - |  |

| | | | | | | | |
|---|---|---|---|---|---|---|---|
| Au₄Er | | 9.8 | L+S | 15/2 | 9.6 | 5 | RKKY | [4] |
| Au₄Tm | | 7.6 | L+S | 6 | 7.6 | 2 | RKKY | [4] |
| Au₄Yb | | 4.5 | L+S | 7/2 | 4.5 | < 6 | RKKY | [4] |
| Au₄Lu | 0 | 0 | - | 0 | 0 | - | - |

## b. Distance-dependent moments and exchange interactions

We compare in Fig. 4a the magnetic moment as a function of nearest neighbor Mn-Mn distance for many ferromagnetic materials where Mn is the only magnetic element. These materials fall into two categories. A small distance of 0.28-0.30 nm leads to itinerant magnetism with direct exchange mechanism where the magnetic moment per Mn varies from 1.5 $\mu_B$ (in Mn$_2$VAl) to 4 $\mu_B$ (MnBi) [24]. Localized magnetism is often found in Mn-based magnets with larger Mn-Mn distances. Magnets where the distance is > 0.4 nm, including many (half) Heusler alloys (and also Au$_4$Mn), all exhibit a moment of about 4 $\mu_B$/Mn.

Strong direct exchange between Mn-pairs can hardly occur in these compounds based on localized 3$d$-states. Previous theoretical studies revealed the crucial role of indirect exchange coupling for the neighbor non-magnetic elements in Heusler alloys with a large Mn-Mn distance. [25,26,27]. Spin polarized $s,p$ bands mediate interatomic covalent interactions between the Mn 3$d$ states. In Au$_4$Mn, we have also found the similar spin polarized 6$s$ and 6$p$ states of Au shown in Fig. 3, so we might think a RKKY interaction is responsible for the ferromagnetism.

We have therefore calculated the interaction strength as a function of Mn-Mn distance $d$ in Au$_4$Mn, shown in Fig. 4b where the points represent the successive shells of Mn neighbors, and $\mathcal{J}_{eff}$ = $Z\mathcal{J}$, where Z is the number of neighbors in a shell. The exchange fluctuates with distance, but there are no antiferromagnetic interactions for the first ten shells with $d \leqslant 1.041$ nm. The exchange interactions are all ferromagnetic or zero. Only beyond that, where that there is little further net contribution to the coupling, the exchange is RKKY-like, oscillating in sign. The results are summarized in Table 2.

The RKKY function $F(\xi) = (\sin\xi - \xi\cos\xi)/\xi^4$ is plotted in Fig 4c, where $\xi = 2k_F d$, $k_F$ is the Fermi wavevector $k_F = (3\pi^2 n)^{1/3}$ for the free-electron gas and $n$ is the number of free electrons per unit volume [1]. In a simple model with one free electron per Au atom, $n = 4.74 \times 10^{28}$ m$^{-3}$ and $k_F$ = $1.12 \times 10^{10}$ m$^{-1}$ and $F(\xi)$ is the dashed curve. In our calculation we find a reduced number of 2.1 free electrons per formula unit due to hybridization of Au 6$s$, 6$p$ orbitals with Mn 3$d$ orbitals, which corresponds to the solid curve. The first- and second- neighbour exchange is of opposite sign in the first case, favoring a spin structure of ferromagnetic chains coupled antiferromagnetically, as in Au$_4$Cr [28]. The calculated magnetic ordering temperature is much lower than the experimental one, assuming $\mathcal{J}_{sd}$ = 0.36 eV for Mn [29] in the RKKY model. In our DFT calculation, the Curie temperature is 460 K if Mn-Au 3$d$-5$d$ interactions are ignored, while it only slightly less, 439 K, if all exchange interactions are considered.

Table 2 Exchange Interactions in Au$_4$Mn by DFT calculation

| Shell Number | Mn-Mn distance (nm) | Z | Z J (meV) | Sum (meV) |
|---|---|---|---|---|
| 1 | 0.404 | 2 | 1.8 | |
| 2 | 0.499 | 8 | 3.4 | |
| 3 | 0.645 | 4 | 3.3 | 8.5 (59%) |
| 4 | 0.760 | 8 | 1.0 | |
| 5 | 0.762 | 8 | 1.3 | |
| 6 | 0.808 | 2 | 0.0 | |
| 7 | 0.912 | 4 | 1.3 | |
| 8 | 0.997 | 8 | 0.0 | |
| 9 | 1.034 | 8 | 1.1 | |
| 10 | 1.041 | 8 | 1.0 | 14.3 (100%) |

We further compare with Mn-based dilute alloy spin glasses where the Mn-Mn distances are even larger and Kondo temperatures are only of order 10 mK. There the dominant exchange is also the s-d RKKY interaction. In fcc **Au**Mn spin glass, a dilute alloy where a few percent of manganese impurities are dispersed in an fcc gold matrix, the effective manganese moment measured in the paramagnetic state is around 5.5 $\mu_B$ [30]. With 1% manganese in this system, the average Mn-Mn distance is 1.19 nm and the spin freezing temperature $T_f$ < 20 K [30]. At these distances, the electrons of Mn 3d shells cannot interact directly and exchange is RKKY via the oscillating spin polarization of the gold conduction band. It is equally likely to be ferromagnetic or antiferromagnetic due to the disorder.

The $T_C$ for localized Mn-based magnets is sensitive to the distance or the number of Mn atoms per volume as shown in Fig. 4d. When the Mn concentration (number of atoms per m$^3$) decreases slightly the $T_C$ drops significantly, as shown from groups of materials with the same valence electrons including NiMnSb-PtMnSb [1], Rh$_2$MnGe-Rh$_2$MnSn-Rh$_2$MnPb [24] and Cu$_2$MnAl-Cu$_2$MnIn-Au$_2$MnAl [24]. This trend is opposite to itinerant magnets MnAs-MnSb-MnBi [1] where $T_C$ increases with atomic distance. Au$_4$Mn is significantly off the trend for itinerant magnets. $T_C$ of Au$_4$Mn is sensitive to the lattice constant. With increasing pressure, both $M_s$ and $T_C$ increase [31] due to the enhanced hybridization with smaller lattice parameters, therefore the magnetism of Au$_4$Mn thin films grown on piezoelectric substrates could easily be modified by strain.

The outstanding property of Au$_4$Mn compared to other 3d magnets with a similarly low concentration of magnetic atoms is its relatively high $T_C$. According to the mean field model, $T_C = 2ZJ\,S(S+1)/3k_B$, where Z is the number of interacting neighbors, $k_B$ is the Boltzmann constant and $J$ is the exchange constant. Taking Z = 10 (counting 1$^{st}$ NN and 2$^{nd}$ NN Mn) and S = 2, we find $J/k_B$ = 9.6 K. Usually itinerant Mn-based ferromagnets have a much larger $J$, due to the close Mn-Mn distance, and smaller moments as shown in Fig. 4e. Other examples with long Mn-Mn distances and large local moments where $J/k_B$ is around 10 K include Heusler X$_2$MnZ and half Heusler XMnZ materials with Z = 12. Others are shown in the localized region in Fig. 4a. If we only account 1$^{st}$ NN Mn and take Z = 2, $J/k_B$ is unreasonably large. Therefore both 1$^{st}$ NN and 2$^{nd}$ NN Mn-Mn interactions contribute to the ferromagnetism on Au$_4$Mn [32].

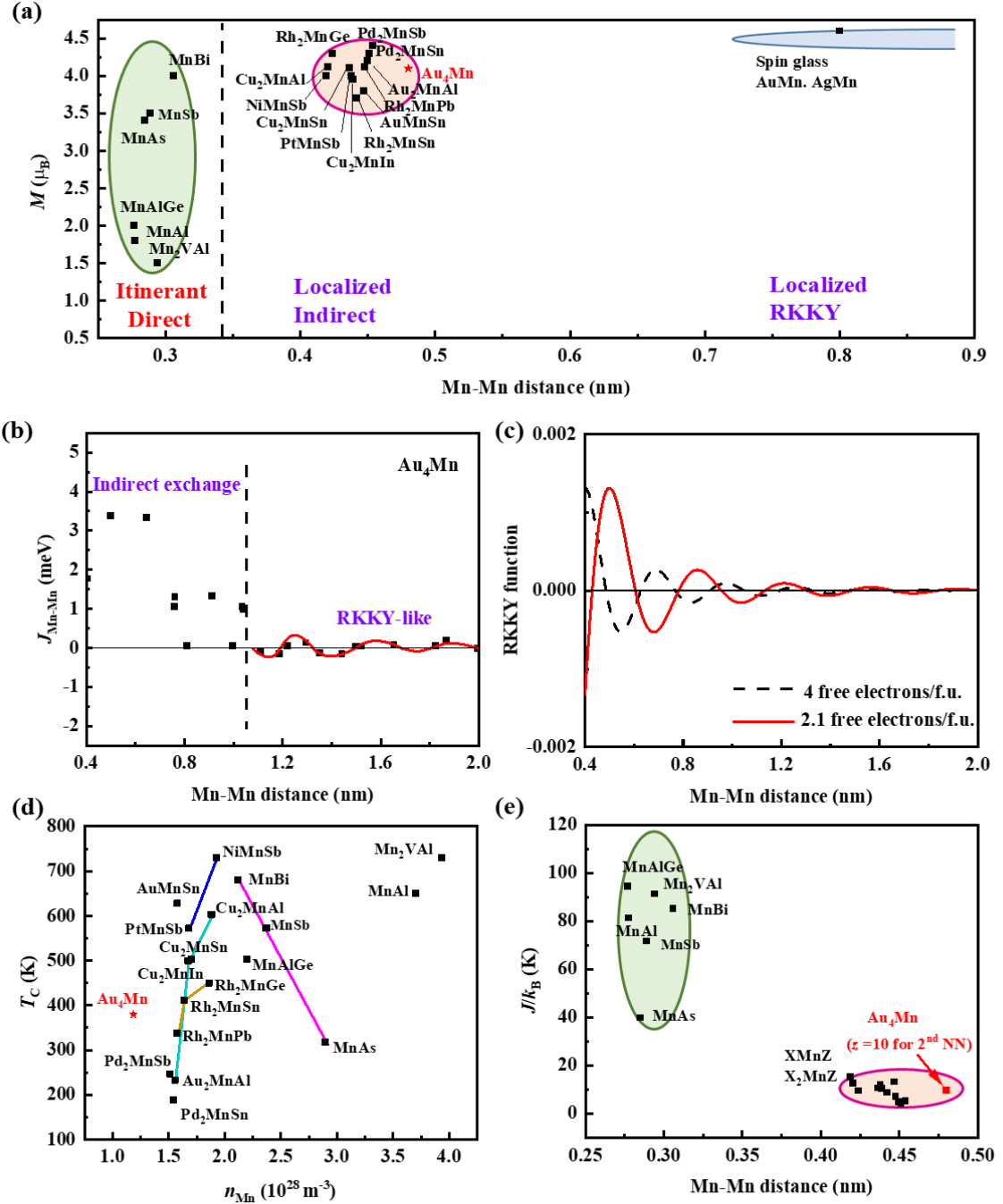

Figure 4. (a) Magnetic moment versus Mn-Mn distance in Mn-based magnets. All materials are ferromagnets where Mn is the only magnetic element, except for the dilute Mn spin glasses, where the Mn moment is deduced from the effective moment in the paramagnetic state [30]. (b) Exchange interaction strength as a function of Mn-Mn distance in $Au_4Mn$. (c) Distance-dependent RKKY function with different free electron numbers. (d) The relationship between $T_C$ and Mn concentration. Data are from Refs. [1,33,34,35,36]. (e) Exchange interaction constant $J$ as a function of Mn-Mn distance in Mn-based alloys. XMnZ and $X_2MnZ$ refer to Heusler and half Heusler alloys as also shown in the localized region in Fig. 4a.

**c. Spin orbit coupling and high field magnetization**

The spin-orbit coupling strength increases with the atomic number. Therefore, the value for Au is larger than most other common elements that can form ferromagnetic alloys [37]. High spin-orbit coupling causes high anisotropy, therefore the gold host significantly enhances the anisotropy.

We now consider possible reasons for the magnetization behavior in high fields. The anisotropy energy $K_1$ is 300 kJm$^{-3}$ at low temperature from torque measurements [2] or 500 kJm$^{-3}$ from our DFT simulations, corresponding to an anisotropy field $2K_1/M_s$ of 1.3-2.1 T. This is significantly smaller than the field of more than 40 T applied in the pulsed field measurement. Therefore, the increase of the moment at high field is not related to a moment that is unsaturated because of polycrystalline magnetocrystalline anisotropy.

The dimensionless high-field susceptibility χ of Au$_4$Mn is of order 10$^{-3}$ (SI), two orders larger than the Pauli susceptibility (10$^{-5}$) and also much larger than that of a ferromagnet such as Fe with χ = 3.8×10$^{-4}$ [38]. This large χ is similar to that of Rh$_2$CoSb of 1×10$^{-3}$ [15], where there is an induced moment 0.3 μ$_B$ of Rh due to the 3$d$-4$d$ hybridization. Therefore, the high field slope in Au$_4$Mn should have a different origin from simple Pauli susceptibility. It is likely to relate to progressive field-induced flipping of a small moment of Au due to the hybridization Mn and Au $d$ orbitals.

However, the first possibility to consider is some manganese disorder that introduces Mn-Mn nearest-neigbour pairs on adjacent sites that couple antiferromagnetically. This results in a decreased net moment and $T_C$ [39]. The additional 0.4 μ$_B$/f.u. under high magnetic field could be attributed to at least 5% of misplaced Mn whose moment flips under magnetic field. We cannot fully rule out some manganese disorder, but it does not explain why a similar phenomenon is observed in the vanadium compound [7]. Such Mn disorder would reduce the Curie temperature, but our samples actually have slightly higher $T_C$ than previously reported [3].

The second possibility is long-range helical or spiral spin structure of Mn. The period would lengthen and the magnetization would eventually saturate in high field. However, neutron diffraction has shown that Au$_4$Mn is a simple ferromagnet [3], and we could find no more stable non-collinear spin structure in our calculation. Again, there is no reason why the vanadium alloy should saturate in the same field.

A third thought, ruled out by the small density of states calculated at the Fermi level, is that Mn atoms increase their moment significantly under applied magnetic field in a paraprocess. The manganese moment does depend on the tetragonal distortion, which is larger at low temperature, as indicated by the anisotropic thermal expansion [40], but our calculation shows that the lattice constant would have to change by much more than 10% to increase the moment from 4.1 to 4.5 μ$_B$/f.u..

Lastly we return to the magnetism of Au. Usually the 5$d$ orbitals of Au are fully occupied and the electrons are unpolarized, so Au is nonmagnetic. But when hybridized with Mn, a small spin moment is induced, which might be field-dependent, like that in the enhanced Pauli paramagnet YCo$_2$ [41]. In fact, an induced spin moment of 0.016 and 0.035 μ$_B$ per Au was observed by XMCD under 3 and 6 T respectively [18, 19], which agrees with our high-field slope of 0.02 μ$_B$/T in Fig. 2a and 2b in this field range. It is also possible that the spin moment on Au rotates continuously from antiparallel to parallel to the manganese moments under applied magnetic field. If, in the ground state, the moments of Au and Mn are antiparallel, flipping a net moment of 0.05 μ$_B$ /Au could explain the experimental result. (0.05×4×2 = 0.4 μ$_B$ /f.u.). A similar argument can be made for the vanadium compound. It seems that the gold in Au$_4$Mn should be regarded as magnetic gold to some degree.

**Conclusion**

    From our study of the magnetism of ferromagnetic Au$_4$Mn, we find that the large, localized manganese moment and significant ferromagnetic exchange coupling at Mn-Mn separations up to 1.04 nm are related to indirect exchange mediated principally by Au 6$p$ electrons. Evidence comes from the well-matched values of $m_s$ and $m_{eff}$, the small Kerr rotation angle and the band structure calculations. The manganese orbital moment is quenched, as in most 3$d$ magnets. The high $T_C$ is a result of the large, localized manganese moments and ferromagnetic exchange in the first ten nearest neighbor shells. Since 80% of its atoms are heavy element Au with a strong spin orbit coupling, its 5$d$ orbital is hybridized with Mn 3$d$ and aligned under high magnetic field. There is a moment of 0.4 $\mu_B$/f.u. on the gold under a magnetic field of 40 T. This hybridization also reduces the Mn moment from an ideal of 5.0 $\mu_B$ to ~ 4.1 $\mu_B$. The uniaxial magnetocrystalline anisotropy at room temperature is manifested by maze surface domains. The magnetic hardness parameter of 1.0 is sufficient to maintain the $c$-axis anisotropy in any shape. We summarized the distance-dependent magnetism of Mn by comparing Au$_4$Mn with many reported Mn-based materials, and find that (1) a small interatomic distance of 0.28 - 0.30 nm leads to itinerant ferromagnetism where the magnetic moment can vary greatly; (2) Localized magnetism with indirect exchange interactions is found in Mn-based magnets with Mn-Mn distances > 0.4 nm, which exhibit a Mn moment of about 4 $\mu_B$.

This work was supported by Science Foundation Ireland, under the MANIAC, SFI-NSF China project (17/NSFC/5294) and ZEMS project 16/IA/4534. We acknowledge the support of the Dresden High Magnetic Field Laboratory (HLD) at HZDR and members of the European Magnetic Field Laboratory (EMFL).